\documentclass[10pt,a4paper,twocolumn]{article}
\usepackage[utf8]{inputenc}
\usepackage[english]{babel}
\usepackage{amsmath}
\usepackage{amsfonts}
\usepackage{amssymb}
\usepackage{graphicx}
\usepackage{url}
\usepackage[numbers,compress]{natbib}
\usepackage{algorithmic}
\usepackage{graphicx}
\usepackage{url}
\usepackage{textcomp}
\usepackage{xcolor}
\usepackage{soul}
\usepackage[all]{xy}
\usepackage{multirow}
\usepackage[margin=0.75in]{geometry}

\author{Jordi Paillisse$^{*}$, Jordi Subira$^{*}$, Albert Lopez$^{*}$, Alberto Rodriguez-Natal$^{\dag}$, \\Vina Ermagan$^{\dag}$, 
Fabio Maino$^{\dag}$ and Albert Cabellos$^{*}$ \\
$^{*}$\textit{\small UPC-BarcelonaTech, Barcelona, Spain} - \textit{\small \{jordip, alopez, acabello\}@ac.upc.edu,} \textit{\small jordi.subira@est.fib.upc.edu}\\
$^{\dag}$\textit{\small Cisco Systems, San Jose, CA, USA - \{natal,vermagan,fmaino\}@cisco.com}}

\title{Distributed Access Control with Blockchain}

\begin{document}
\maketitle
\begin{abstract}
The specification and enforcement of network-wide policies in a single administrative domain is common in today's networks and considered as already resolved. However, this is not the case for multi-administrative domains, e.g. among different enterprises. In such situation, new problems arise that challenge classical solutions such as PKIs, which suffer from scalability and granularity concerns. In this paper, we present an extension to Group-Based Policy -a widely used network policy language- for the aforementioned scenario. To do so, we take advantage of a permissioned blockchain implementation (Hyperledger Fabric) to distribute access control policies in a secure and auditable manner, preserving at the same time the independence of each organization. Network administrators specify polices that are rendered into blockchain transactions. A LISP control plane (RFC 6830) allows routers performing the access control to query the blockchain for authorizations. We have implemented an end-to-end experimental prototype and evaluated it in terms of scalability and network latency. 
\end{abstract}

\section{Introduction}
Group-Based Policy (GBP), or policy-based networking \cite{Verma2002} is a declarative approach to defining  network behaviour. Network administrators specify network endpoints, groups of endpoints and their policies using a high level language, which is later translated to network configurations. One of these languages, GBP is widely employed in the industry, for example in OpenStack's Neutron network API \cite{openstack18}. It can define rules between servers and clients, service chains, etc. 

Until now, GBP has been conceived as a language for a single administrative domain. In this paper, we analyze if we can extend it to several administrative domains, preserving at the same time their independence. For example, we want to make it possible for an administrator in company B to allow a VPN connection from a user in company A by simply typing:

\texttt{\footnotesize{createPolicy from=userA to=VPNserverB action=allow}}

%A common problem in today's networks is trust management between different administrative domains. In other words, authorizing several users access to a resource from a different trust domain. Examples of this situation are the establishment of a VPN from one company to another, or allowing certain computers access to a network of IoT devices. 

Typically, there are two solutions for this scenario: (i) manual, by means of issuing a digital certificate and giving it to the users (so they use it later to authenticate the connection), or (ii) leveraging structures based on PKI systems, namely cross-domain certification or bridge CA certificates \cite{josang2013pki}. These structures allow the co-existence of several CAs and ensure mutual trust.

However, these approaches present some limitations that have hindered their deployment. First of all, scalability: in scenarios with thousands or tens of thousands of users, the manual approach is unfeasible, and cross-certificating $N$ domains means -in the worst case- issuing $\sim$ $N^{2}/2$ certificates \cite{josang2013pki,slagell06}. Second, granularity: it is not possible to define different policies for different users without issuing more certificates, further affecting scalability. Finally, management: PKIs are cumbersome to manage, especially day-to-day operations like adding and removing users, revocation (requires a CRL subsystem) or key rollover.

%Finally, trust: at the end of the day, both \hl{cross-domain certification} \textcolor{red}{remove due to footnote?} and bridge certificates schemes blindly trust the CA, which means that resource holders cannot \emph{retain} control over their resources, it is rather the CA that has the final say\footnote{It should be noted that this situation could be avoided in a cross-certificated PKI by carefully selecting (revoking) the trusted (untrusted) certificates, but this architecture scales poorly and does not offer user-level granularity.}. In a cross-enterprise scenario, this is usually unacceptable: a company has to be able to unilaterally revoke access to a particular resource if it judges appropriate.

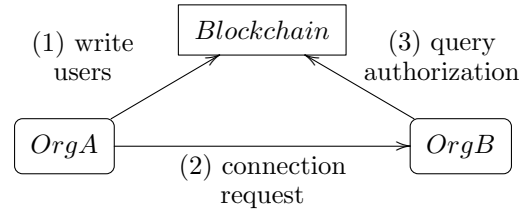
\begin{figure}[tbp]
\begin{displaymath}
\xymatrix{   & *++[F]{Blockchain} &  \\ 
             *++[F-:<3pt>]{Org A} \ar[ur]^{\txt<8pc>{(1) write \\users}} \ar[rr]_{\txt<8pc>{(2) connection request}} & &  *++[F-:<3pt>]{Org B} *\ar[ul]_{\txt<8pc>{(3) query \\ authorization}}
}
\end{displaymath}
\caption{Global architecture.}
\label{fig:global}
\end{figure}

With these limitations in mind, in this paper we propose using a blockchain to overcome them. In such blockchain, each organization defines its users and resources, and specifies which users -from other organizations- can access its resources. Upon an access request, routers query the blockchain to verify authorization (fig. \ref{fig:global}).

Thanks to blockchain's particular properties, we can design an access control system that improves on several of PKI's limitations: (i) \emph{increased scalability: }when we establish a new relationship in a PKI, we have to cross-certificate the new entity with the rest. In a blockchain, however, we can directly reference previous transactions/users. This reduces the number of required certificates (transactions in this case). (ii) \emph{improved granularity and flexibility}: since we can associate each resource or user with a private key, we can alter its state without affecting the rest. This includes both its validity and other data, for example, we can assign different policies to different users. 
%\hl{REMOVE? In a PKI, a revocation affects all certificates below the CA, not only a specific one, i.e. there is no partial revocation. -Unsure: you can revoke certificates individually-} 
(iii) \emph{simpler management:} the transactional nature of blockchain makes management simpler: the aforementioned common operations (key rollover, revocation) can be encoded as new transactions, instead of requiring a dedicated subsystem, like CRLs and manifests.

%(i) \emph{decentralized:} each organization retains control over its resources (they are associated with a private key), so it can revoke access at any time. In contrast, in a PKI this translates to a request to the CA. 

In this paper we present an architecture to support this use-case, we describe a practical end-to-end implementation and evaluate its performance to demonstrate its feasibility. Our results show that we can store thousands of access polices with modest storage, and achieve linear update times on a permissioned blockchain.

\section{Why Blockchain?}
Our use-case presents two particular characteristics: (i) its participants have limited trust in each other, and (ii) they want to retain full control over the access policies. This is because in a multi-enterprise scenario: (i) companies are not willing to leave access control to a third party, and (ii) each company must be able to revoke any access policy at any moment in time, respectively. 

These requirements match the characteristics of any blockchain. Regarding the first demand, its consensus algorithm ensures that no single entity controls the blockchain and avoids having to fully trust its participants. The second requirement is covered by the fact that blockchain assets are controlled by their associated private key owner, not by a centralized entity.

On the other hand, an approach like this is much more complex in a classical PKI because it cannot fulfill the previous two requirements. The first one because the CA is the single point of trust in the system, which forces all participants to trust it. Furthermore, it cannot meet the second as a consequence of the centralized trust: the CA can unilaterally alter state by means of certificate revocation. 

%This is due to the fact that the CA is the single point of trust in the system, thereby forcing all participants to trust it and accept its decisions.

%Explain why a classical pki cannot do this:
%1st req: CA is a single point of trust, has to be trusted
%2nd req: CA can revoke any downstream certificate

%Explain why bridge CA cannot (same problem) and cross-domain has limited scalability
As mentioned before, other PKI schemes could provide equivalent functionality, such as bridge CA certificates, but in this case the bridge CA certificate is still a single point of trust, thus it is not significantly different from a conventional CA. Cross-domain certification may prove useful, however, it presents scalability limitations because each new CA has to cross-certify with all the existing ones.

In addition, a blockchain can alleviate this scalability concerns: we can reduce the number of required certificates (transactions in this case), since a blockchain allows directly referencing existing transactions, instead of re-certifying with the PKI CA. In turn, this simplifies the verification of the chain of authorizations, i.e. it is not necessary to go up the CA, then down to the cross-certificated one. Authorization emerges directly from the originating organization.

Finally, thanks to emerging private blockchain platforms we can provide a certain degree of privacy for their users (as opposed to public blockchains like Ethereum) and improve some of their performance metrics (sec. \ref{sec:arch:bc}).

\section{Architecture}
%L’estructura general (FIG) (es pot posar en termes de PDP, PEP, que vas trobar):

We can describe our architecture as a three-layer system (fig. \ref{fig:layer}):
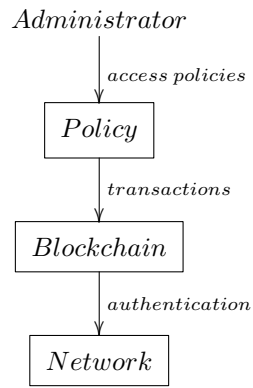
\begin{figure}[!tbp]
\begin{displaymath}
\xymatrix{   Administrator \ar[d]^{access\:policies} \\
              *++[F]{Policy} \ar[d]^{transactions}  \\
              *++[F]{Blockchain} \ar[d]^{authentication}\\
              *++[F]{Network}
}
\end{displaymath}
\caption{Layered architecture.}
\label{fig:layer}
\end{figure}

\emph{\textbf{Policy: }}an intent-driven interface allows administrators to specify users, resources and access policies. These polices are rendered into blockchain transactions.

\emph{\textbf{Blockchain: }}a blockchain stores all the information and ensures its integrity and accuracy.

\emph{\textbf{Network: }}routers access the blockchain via an API to determine if a particular user can access a specific resource. If the user is authorized, they retrieve authentication information to establish a security association and allow the connection.

The following sections provide details on each element.

\subsection{Policy interface}\label{sec:interface}
Administrators use a simple CLI, based on GBP, to perform management operations, such as creating/deleting users, groups of users, policies and resources, as well as querying the blockchain for specific policies, users, etc. We have chosen GBP because it is widely adopted in the industry \cite{openstack18} and its semantics align pretty well with our use-case.

Specifically, we can accommodate our use case to the OpenStack syntax simply re-using some of its commands. For example, consider that organization B wants to grant access to its internal database to Alice from the Human Resources department of an external organization A (figure \ref{fig:example}). First, company A creates a member for Alice:
\begin{figure}[!tbp]
\centerline{\includegraphics[width=\columnwidth, trim={1cm, 7cm, 0.5cm, 2.5cm}, clip]{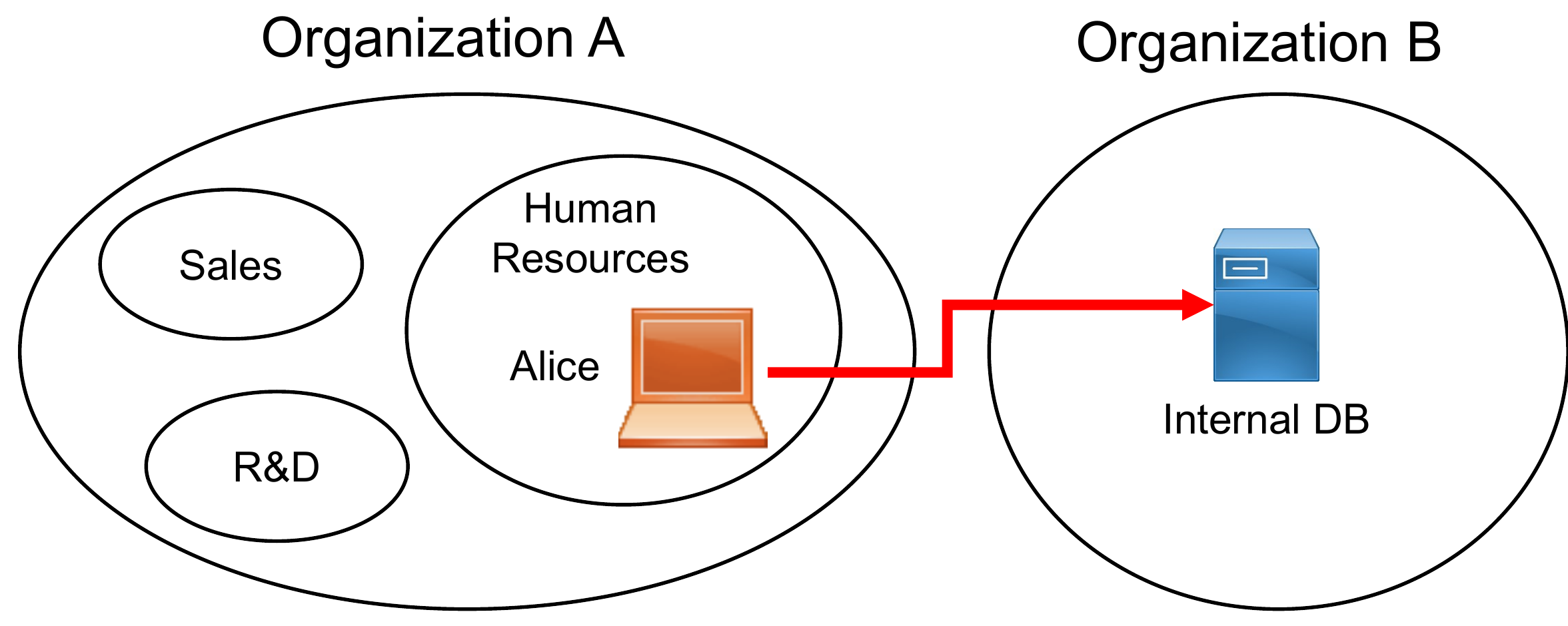}}
\caption{Example scenario}
\label{fig:example}

\end{figure}

\texttt{\footnotesize{gbp member-create alice}}

Then, company B creates a group for company A's user and adds Alice into it:

\texttt{\footnotesize{gbp group-create dbaccess --add:orga.alice}}

It also creates the internal database as a member:

\texttt{\footnotesize{gbp member-create internalDB}}

Finally, organization B creates the policy associated to its database and company A's user, allowing access from the group \emph{dbaccess} to its member \emph{internalDB}:

\texttt{\footnotesize{gbp policy-rule-create external-human-res --src:dbaccess --dst:internalDB  --actions allow}}

The GBP syntax can be extended with more options, for example, adding a one week timeout to Alice's membership in the \emph{internalDB} group:

\texttt{\footnotesize{gbp group-create dbaccess --add:orga.alice --timeout 1w}}

This custom logic can be easily implemented thanks to the ability of some blockchains to run smart contracts. 
\subsection{Blockchain}\label{sec:arch:bc}
In this section we discuss two major design decisions we took for our blockchain.

\emph{\textbf{Participants:}}
We believe that a private blockchain (only authorized members can access it) fits better in this scenario than a public, mainly because its participants are not willing to make their access policies public. Communicating access policies only to a group of companies is sufficient for correct operation. 

\emph{\textbf{Consensus algorithm:}}
We argue that a BFT protocol suits our use case, due to the following reasons: (i) Security: classical BFT protocols such as XFT \cite{liu2016xft} or BFT-SMART \cite{Bessani14} offer proven security guarantees, as opposed to PoW or PoS algorithms, some of which lack a formal security analysis or a mature implementation. (ii) the access-control PKI of the private chain can be re-used in the BFT protocol, since they require some kind of node authentication. (iii) Higher throughput: BFT algorithms typically reach consensus faster than PoW or PoS, thus increasing the amount of transactions per second. (iv) Immediate finality: in a BFT protocol, when a transaction has been added in the chain, it will never be removed. On the contrary, a Bitcoin fork prevents immediate finality, and (v) We can avoid well-known PoW/PoS drawbacks, e.g. high energy consumption or limited throughput.

%In addition, the two latter algorithms are tailored for public blockchains, which makes them less suitable in a private chain. 
%In addition, private blockchains present higher transaction rates and immediate finality than their public counterparts. 

Finally, it should be noted that BFT-based chains suffer from scalability concerns, i.e., they cannot scale to as many users as well-know PoW or PoS chains like Bitcoin (in the order of millions). However, this is not our case: a chain with hundreds or even tenths of companies would be perfectly functional.

% perquè HL
% Perquè vam triar HL: private, molta documentació, projecte sòlid, business-oriented
% chaincode flexible
% msp retain control

\subsection{Network}\label{sec:network}
In order to perform the access control, we have chosen the Locator/ID Separation Protocol (LISP, RFC 6830). LISP is a request-response protocol that allows the communication of control and data planes. In our scenario, routers can easily retrieve the blockchain access control policies from the control plane with minimal modification of the base protocol. For this particular use-case, LISP is conceptually equivalent to OpenFlow \cite{McKeown2008}. Hence, we can use other protocols for this task, such as the aforementioned OpenFlow or P4 Runtime \cite{p4r18}.

In a nutshell, we store the access policies in the LISP control plane and update them through the blockchain. LISP-enabled routers query the control plane to determine if a particular user can access the requested resource. Users authenticate to the router by means of including their signature in the LISP control plane messages (Map Request and Map Reply).

\subsection{Typical Workflow}\label{sec:wf}

\begin{figure}[tbp]
\centerline{\includegraphics[width=\columnwidth, trim={0, 1.5cm, 0, 1cm}, clip]{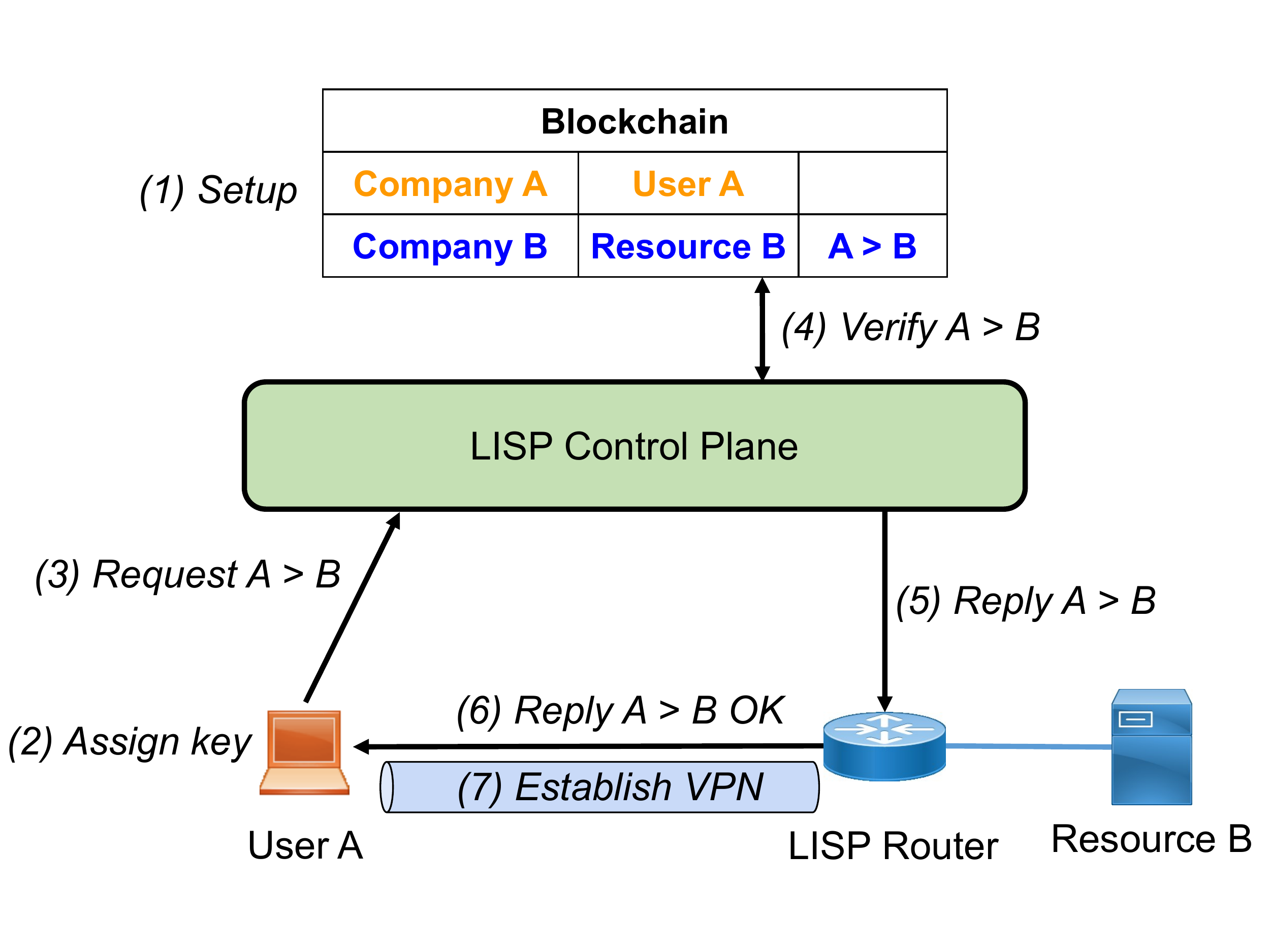}}
\caption{Typical architecture workflow.}
\label{fig:workflow}
\end{figure}

Fig. \ref{fig:workflow} presents an example of the typical workflow in this architecture, in which two companies set up a secure connection from User A of company A to Resource B located in company B.
\begin{enumerate}
\item At setup time, administrators from both companies store the required information in the chain. Company A adds User A and its public key. Company B details its resource (Resource B) and grants access to company A's user (A $>$ B). They use a CLI similar  to the example in sec. \ref{sec:interface}.
\item Company A assigns User A its credentials (public-private keypair), with the public key being the one in the blockchain.
\item When User A wants to connect to Resource B, it sends a LISP control message to the LISP Control Plane. The message is signed by User A. 
\item  The LISP Control Plane verifies the signature and checks the access policy against the blockchain.
\item If they are correct, it sends a reply message to Company B's LISP Router.
\item The LISP router sends a reply message with the cryptorgraphic material for data plane encryption, in order to establish a security association with User A. This message is encrypted with User A's public key.
\item The LISP router and User A start a secure connection, e.g. with LISP-CRYPTO (RFC 8061) or other VPN protocols (L3VPN or equivalent).
\end{enumerate}

%\hl{specifiy security assumptions? xTRB $>$ EIDB secure (inside company B),  and MS $>$ XTRB secure (inside company B)?, no xTR authentication, only user?}

\section{Implementation}
We have built an end-to-end prototype encompassing the three components of the aforementioned architecture: GBP interface, Blockchain, and Network. It is available as open-source code\footnote{https://github.com/JordiSubira/DGBP}.

\subsection{GBP Command Line}
We designed a CLI inspired on GBP commands that allows the creation/deletion/retrieval of users, groups of users, resources and access policies, highly similar to the examples in sec. \ref{sec:interface}. For example \texttt{create resource} creates a new resource for the organization. Additional options in the commands specify the IP address of the resource, the public key of a user, etc.

\subsection{Blockchain}
Our prototype is based on the Hyperledger project (HL), an open-source permissioned blockchain implementation. Specifically, we have chosen its Fabric \cite{fabric18} framework because of its maturity, flexibility and business-orientation. Moreover, thanks to Fabrics's channels, we can establish private communications among sub-sets of companies if privacy  is a strong concern. In this section we summarize the different configuration parameters and implementation details of our prototype in the HL Fabric framework.
\subsubsection{Assets}We defined the following elements in the chain:

\emph{\textbf{Users:}} Source endpoints, identified by a public key and including other information: originating organization, IP address, name and department.

\emph{\textbf{Departments:}} A group of users within an organization, identified by department name and their belonging organization\footnote{They identify groups of users in order to simultaneously create  policies for several users.}.

\emph{\textbf{Resources:}} Destination endpoints, identified by an IP address and the associated organization.

\emph{\textbf{Policies:}} Access control lists that grant access either from a source endpoint (user) or from a group of users (department) to a destination endpoint (resource). They are identified by a composite key (source-destination). Typically, the source endpoint is a user or department of another organization and the destination is a resource of the issuing organization. Policies can contain other information, such as the time frame in  which the connection is allowed or an expiry time. 	

\subsubsection{Membership Service Provider}Each organization is identified by a MSP (Fabric's PKI for blockchain nodes).

%It contains all necessary cryptographic material and lists the identity of their components (peers, orderers, clients, and so on). 

%used by an organization to define its own trust domain. The MSP 
%Thus, this provides enough information to ensure authentication in every process if HyperLedger Fabric. 

%It is general to any HL implementation, right?
%In our scenario, every organitzation must ensure that all crypto material is correctly distributed among their nodes. Likewise, organitzations must promise a loyal and secure internal behaviour. Cryptographic information of every MSP's component is collected and placed into Genesis Block at bootstrap time. 

\subsubsection{Chaincode}We imposed as a global constraint that only the organization that creates an asset can alter its state (e.g. delete, associate with another asset, etc). We enforce this by binding all assets to their respective MSP, and rejecting any modification from a non-owner MSP.
%Hyperledgers's realization of Smart Contracts is known as Chaincode and allows the definition of the state transition rules that implement the business logic. 

In addition, any organization within the same HL channel can query information about any asset of any organization in the chain.

%we defined the following rules. Each organization can insert and remove:
%HL Fabric employs chaincode as Smart Contract. This software, which contains the business logic, is distributed among the endorsers within HL network. Each one of those endorsers executes the chaincode, so as to process transactions and checks their validity. 
%\begin{itemize}
%\item Users. Users are binded to their creating organization MSP. Only this organization is can remove them from State DB.
%\item Departments. Departments are also binded to their creating MSP. Only this organization is enabled to remove their own departments from State DB.
%\item Resources, which identify destination endpoints. These are straightaway binded with the organization's MSP and only this organization is allowed to remove their own resources from HL.
%\item Access policies. Each organization creates an access policy compulsory defining an own resource as destination and any user or group of users as source. Additionally, other complementary fields can be defined such as range of days or hours on which connection is granted or validity deadline, among others.  Only the issuer organization is allowed to remove their own policies from State DB.

%\end{itemize}

\subsubsection{Endorsement policy}In our implementation, all members have to endorse any transaction. However, other schemes are possible thanks to Fabric's flexibility. Depending on the level of trust among the participating organizations and the particular use-case, we can adjust the minimum number of endorsements. Some examples are: half + 1 of the members, $2f + 1$ valid signatures out of $n$ endorsers (assuming $f$ faulty endorsers and $n>3f$), or AND/OR syntax (\texttt{member A OR members (B,C,D)}), etc. 
% There is an inherent tradeoff between flexibility and security
%It is worth to mention that this policy could be changed depending on application flexibility and security requirements.

%particularities of the scernario (nivell de trust que tenen entre elles les empreses, o 
%NO>>> no diferencia >>>es podria demanar endorsement difereent segon el tipus de transaccio?)
%No ho deixen fer

\subsubsection{Ordering Service}We leveraged the SOLO ordering service (i.e. a centralized orderer), so we could ease development. However, in a production setup Apache Kafka could be a good fit, because it can tolerate several faulty or disconnected nodes.

In scenarios with low trust among participants, BFT ordering services can be easily plugged thanks to Fabric's modular design. However, we believe that a CFT algorithm is enough for this use-case since a double-spend does not make sense here\footnote{n.b. in a CFT environment some attacks, such as censoring a transaction, may become feasible}. In addition, HL's endorse-order-validate transaction lifecycle offers a variety of mechanisms to prevent or detect misbehavior.

%Depends on use case and assumed level of securety

%Owing to our business case and the new approach provided by the three stage transaction lifecycle, consensus is ensured. In HL fabric, consensus comprises more than just an agreement in the order of transactions, but several autorship authentications and verification of correctness. 

\subsection{OpenOverlayRouter Software Router}
In order to effectively perform access control, we took advantage of an open-source LISP implementation, Open Overlay Router (OOR \cite{rodriguez2017programmable}). We made a slight modification to its Tunnel Router mode: when it receives a Map Request packet, it queries Hyperledger with the source and destination endpoints, via an ad-hoc API. Hyperledger checks if the pair of (source, destination) is allowed to establish a connection and notifies OOR. If they can connect, OOR then responds to the source with a Map Reply message, otherwise takes no action. This way, unauthorized users do not receive a response in the form of a Map Reply and do not know where to connect. Of course, a production setup requires additional security mechanisms, as outlined in section \ref{sec:network}.

%checks HL Fabric in order to grant access, as long as policies were valid.
%Parlar de la necessitat de nous missatges a LISP. Com poses al foot-note
%Parlar del contingut d'aquests missatges, al menys per deixar clar la información que s'utilitza.

\section{Experimental Evaluation}
\subsection{Scenario}
We set up an experimental scenario on a PC running Ubuntu 16.06 and a quad-core Intel i5 CPU 650 @ 3.20GHz. Table \ref{tab:setup} summarizes HL parameters during the experiment. Thanks to the Docker containerization of HL, we could emulate 4 organizations, each with 2 peers, all in the same PC. We artificially generated around 1 million policies and users to evaluate the read latency, and added at most 15 endorsers to estimate the write latency.
%These are default configurations unless noted otherwise

\begin{table}[tbp]
\caption{Experimental Setup}
\begin{center}
\begin{tabular}{|c|c|}
\hline
Number of organizations& 4\\
\hline
Typical key + value size & 32 bytes\\
\hline
 Number of channels&  1\\
\hline
State database & LevelDB key-value store\\
\hline
\multirow{2}{*}{Endorsement Policy} & AND (Org1,Org2,\\
& Org3,Org4)\\ 
\hline
Ordering Service &SOLO \\
\hline
Block timeout & 100 ms\\
\hline
\end{tabular}
\label{tab:setup}
\end{center}
\end{table}

\subsection{Results}
We carried out several experiments on our implementation to characterize its performance and have an understanding of its scalability. 

\begin{figure}[tbp]
\centerline{\includegraphics[width=\columnwidth]{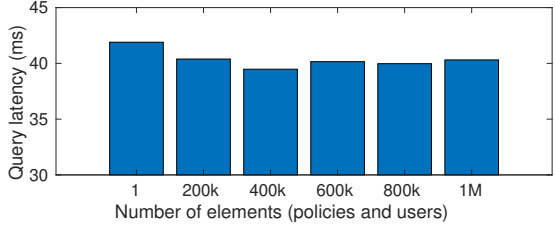}}
\caption{Hyperledger CouchDB query latency.}
\label{fig:query}
\end{figure}

\textbf{\emph{Read latency: }}Fig. \ref{fig:query} presents the average query time for different number of elements in the chain. In this case the state DB was CouchDB\footnote{HL allows using both LevelDB and Couch DB as state DB, with similar performance.}. We can see that it revolves around 40 ms regardless of the number of elements, because Couch DB is a key-value store (these type of databases present constant query latency). It should be noted that the query performs exact matches of pairs of source and destination IP addresses, and that future work should also support longest-prefix matching.

%Jordi used Level DB because bootstrap was faster when doing the tests. For the read latency he kept the original couch DB which supports enriched queries

\begin{figure}[tbp]
\centerline{\includegraphics[width=\columnwidth]{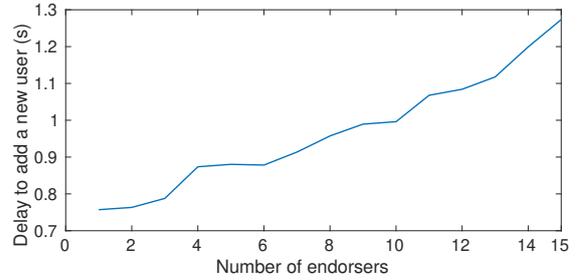}}
\caption{Latency to add a new user for different number of endorsers.}
\label{fig:adduser}
\end{figure}

\textbf{\emph{Write latency: }}Fig. \ref{fig:adduser} plots the time required to add a new user depending on the number of endorsers in the network. As we can see, the latency grows linearly with the number of endorsers, because each new endorser is an additional signature that the issuer has to verify (the current HL implementation makes this verification sequentially). This result is in line with in a recent benchmark of the HL platform \cite{thakkar2018performance}.

%RTT delay: quan trigo a enviar dades LISP des que començo la connexio (primer MReq) vs. num empreses o recursos o usuaris (TBD). 
%Chain delay: quan trigo a saber si un usuari esta autoritzat (ie delay de la consulta a l’API que has fet) vs. num empreses o recursos o usuaris (TBD)
%Setup delay: quan es triga a afegir un nou usuari / recurs a la chain
%Potser al final caldra mesurar bootstrap time i chain size (no és critic en aquest cas i és facil de calcular)

\begin{figure}[tbp]
\centerline{\includegraphics[width=\columnwidth]{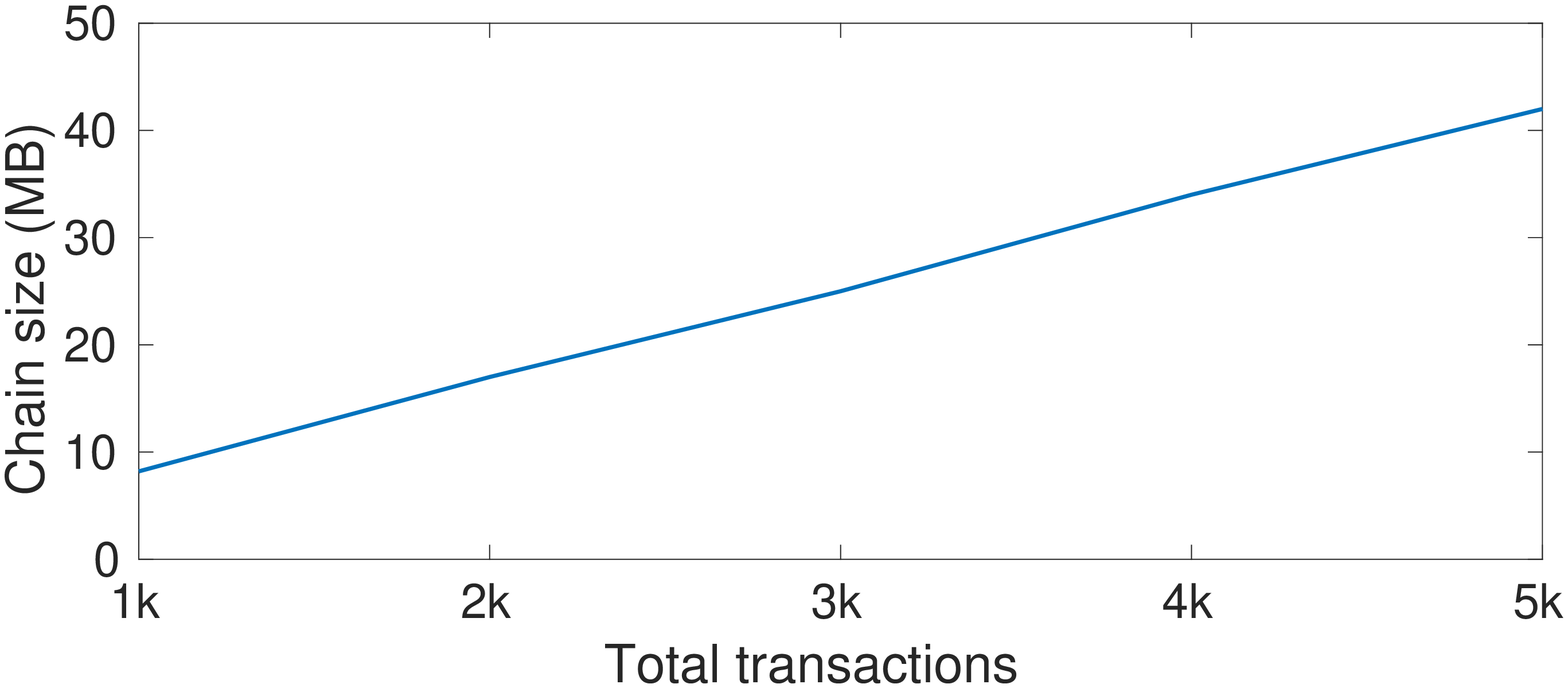}}
\caption{Chain size vs. number of transactions, in a setup with four endorsers. }
\label{fig:sizetx}
\end{figure}

\begin{figure}[tbp]
\centerline{\includegraphics[width=\columnwidth]{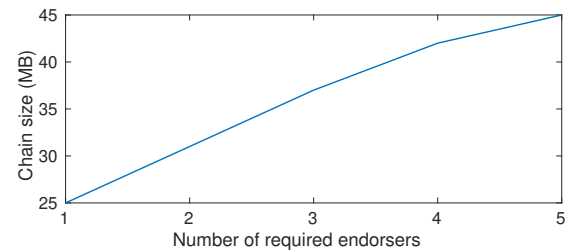}}
\caption{Chain size depending on the number of required endorsers (5k transactions in the chain).}
\label{fig:sizeendorsers}
\end{figure}

\textbf{\emph{Chain size: }}We were also interested in the chain size, i.e. required storage. Figs. \ref{fig:sizetx} and \ref{fig:sizeendorsers} show the total chain size depending on the number of transactions and endorsers, respectively. As expected, in both cases the size grows linearly with the number of transactions or endorsers (in the latter case because more endorsers mean more signatures per transaction). We can see that these situations require very modest storage. Thus, we can safely assume that scenarios with a considerable amount of participants (e.g. 1k endorsers would demand $\sim$25 GB) or a long transaction history (1M transactions take up $\sim$10 GB) can be easily supported.

\begin{figure}[tbp]
\centerline{\includegraphics[width=\columnwidth,trim={1.5cm, 0, 2.5cm, 1.5cm}, clip]{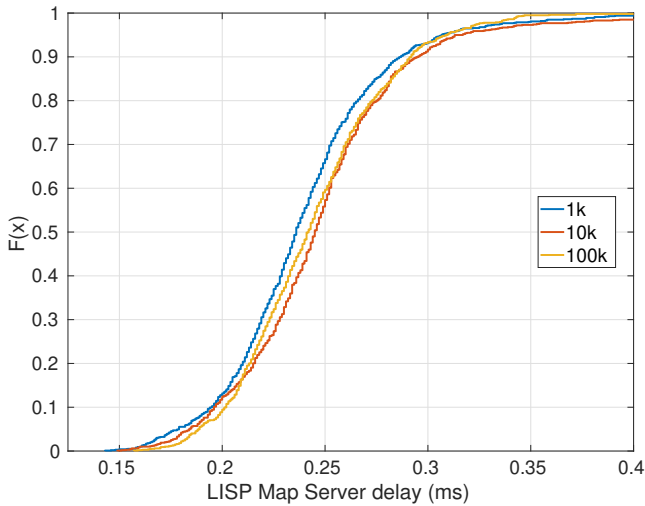}}
\caption{CDF for LISP Map Server query delay for 1k, 10k or 100k pairs of source, destination nodes.}
\label{fig:lisprtt}
\end{figure}

\textbf{\emph{Network latency: }}Fig. \ref{fig:lisprtt} presents the query time CDF of a LISP control plane node (Map Server) storing 1k, 10k or 100k pairs of source, destination pairs. In other words, given a source node, how long does it take to find the authorized destination node(s). Since this test was performed in a local network, we consider the communication delay  negligible.
%\hl{It is not EXACTLY like this but it is equivalent}

We can see that the majority of the queries are completed in less than 0.35 ms, roughly two orders of magnitude below the HL database delay. This is mainly due to two reasons: (i) the Map Server is implemented in C (whereas the queries in fig. \ref{fig:query} go through HL's Node.js API, CouchDB and back) and (ii) data is stored in a Patricia Trie, a tree optimized for prefix queries. In addition, the delay is independent of the number of elements thanks again to the Patricia Trie: the delay depends on the length of the elements (source endpoints, IP addresses in our implementation), not the number. 

%\textbf{\emph{Total delay: }} \hl{Add RTT LISP + query latency+  communication delay? but the lisp is very low :(}
\subsection{Discussion}

The previous results demonstrate that the proposed system can easily scale to meet the demands of a federation of several organizations. Table \ref{tab:summary} outlines several metrics and their requirements in terms of scalability.

On one hand, both the read and network latencies can support high query rates. The read latency presents a constant response time regardless of the number of elements in the chain (in case of exact matches), and the network server storing the access policies, linear with the length of identifiers. 

On the other hand, the write latency can suffer if we have a large amount organizations in the same blockchain. However, it is not as critical as the read latency because we can tolerate a delay up to several of seconds when adding a new user.

Finally, the chain size obviously depends linearly on the number of transactions, but also on the number of endorsers. This last relationship puts an additional strain on scalability, because it affects both chain size and write latency. There is a tradeoff here between a small number of endorsers (small write latency and chain size, but more centralized trust in a narrow set of participants), and a large number of them (higher write latency and size but more distributed trust). Thus, special consideration should be put in the number of endorsers and the endorsement policy to achieve an equilibrium between a tolerable write latency and the expected number of endorsements.  

\begin{table}[tbp]
\caption{Scalability Analysis}
\begin{center}
\begin{tabular}{|c|c|}
\hline
Read latency    & \multirow{2}{*}{constant}  \\
(exact match) & \\
\hline
Write latency  & linear w.r.t. number of org.\\
\hline
\multirow{2}{*}{Chain size} &  linear w.r.t. number of\\
 & transactions and endorsers\\
\hline
\multirow{2}{*}{Network latency}  & linear w.r.t. identifier size\\
&+ propagation delay \\
\hline
\end{tabular}
\label{tab:summary}
\end{center}
\vspace{-0.75cm}
\end{table}

\section{Related Work}
%There are already several proposals in the literature \cite{Bozic2016} that leverage blockchains for a wide range of network applications: naming systems \cite{ali2016blockstack}, IP addresses \cite{paillisse18}, BGP announcements \cite{delarocha17}, Information-Centric Networking \cite{Fotiou2016}, mesh networks \cite{selimi2018towards} or IoT \cite{Christidis2016}.

There are already several proposals in the literature \cite{Bozic2016} that leverage blockchains for a wide range of network applications, such as mesh networks \cite{selimi2018towards}, IP addresses \cite{paillisse18}, etc.

The most closely related work to ours is \cite{francesco17}, which implements Attribute-Based Access Control (ABAC) policies over the Bitcoin blockchain. It presents three main differences with respect to our work: (i) it focuses on access control for individual users, unlike our organization-based approach, (ii) it allows transferring access control rights between users, and (iii) does not consider using a private chain or a different consensus algorithm. Hadi \cite{hashemi2017decentralized} proposes a data distribution system in which the blockchain is the data persistence layer, but is also user-centric and more oriented towards data storage and messaging services rather than networking.

Finally, there is also a growing body of work on blockchain-based access control for IoT: \cite{iotchain18} leverages a blockchain to store access permissions for IoT devices with a strong emphasis on key management and distribution. \cite{pinno18controlchain} also provides authentication, authorization and auditing for IoT but separates them in four independent blockchains, and is generic enough to support a wide range of access control models typical of IoT, while in this paper we concentrate on a specific language, GBP.

%In addition, \cite{bui2016application} presents an interesting perspective on revocation management in distributed access control with blockchains that aligns well with our ideas.

%and distributed cloud storage \cite{wang18storageblockchainfinegrain,sukhodolskiy18cloudaccess}.

\section{Conclusion}
In this paper we have presented, implemented and evaluated an architecture to support access control in cross-domain communications. In order to reduce the burden on network administrators, the front-end builds on GBP, a well-known intent-driven language. A permissioned blockchain distributes network polices, and helps overcome drawbacks of conventional solutions while at the same time maintains the independence of each organization. Our experimental evaluation shows that this design can easily scale to -at least- tenths of organizations with modest storage requirements.

\section*{Acknowledgment}
This work has been supported by the Spanish MINECO under contract TEC2017-90034-C2-1-R (ALLIANCE) and the Catalan Institution for Research and Advanced Studies (ICREA).

%Put sponsor acknowledgments in the unnumbered footnote on the first page.
%Cisco?

%\section*{References}

%Please number citations consecutively within brackets \cite{b1}. The 
%sentence punctuation follows the bracket \cite{b2}. Refer simply to the reference 
%number, as in \cite{b3}---do not use ``Ref. \cite{b3}'' or ``reference \cite{b3}'' except at 
%the beginning of a sentence: ``Reference \cite{b3} was the first $\ldots$''
%
%Number footnotes separately in superscripts. Place the actual footnote at 
%the bottom of the column in which it was cited. Do not put footnotes in the 
%abstract or reference list. Use letters for table footnotes.
%
%Unless there are six authors or more give all authors' names; do not use 
%``et al.''. Papers that have not been published, even if they have been 
%submitted for publication, should be cited as ``unpublished'' \cite{b4}. Papers 
%that have been accepted for publication should be cited as ``in press'' \cite{b5}. 
%Capitalize only the first word in a paper title, except for proper nouns and 
%element symbols.
%
%For papers published in translation journals, please give the English 
%citation first, followed by the original foreign-language citation \cite{b6}.

\bibliographystyle{unsrtnat}
\bibliography{IEEEabrv,gbpbiblio}

\end{document}